\documentclass{aa}  
\usepackage{graphicx}
\usepackage{txfonts}
\begin{document}
%
   \title{Determination of the mass function of extra-galactic GMCs via NIR color maps}

   \subtitle{Testing the method in a disk-like geometry}

   \author{J. Kainulainen \inst{1 \and 2} \and M. Juvela\inst{2} \and J. Alves \inst{3}  
          }

   \offprints{J. Kainulainen}

   \institute{European Southern Observatory, Karl-Schwarzschild-Str. 2, D-85748 Garching bei M\"unchen\\
              \email{jkainula@astro.helsinki.fi}
   \and Observatory, P.O. Box 14, FI-00014 University of Helsinki 
   \and Calar Alto Observatory, Centro Astronomico Hispano, Alem\'an, C/q Jes\'us Durb\'an Rem\'on 2-2, 04004 Almeria, Spain  
}

   \date{<>; <>}

 
  \abstract
   {The giant molecular clouds (GMCs) of external galaxies can be mapped with sub-arcsecond resolution using multiband observations in the near-infrared. However, the interpretation of the observed reddening and attenuation of light, and their transformation into physical quantities, is greatly hampered by the effects arising from the unknown geometry and the scattering of light by dust particles.}
   {We examine the relation between the observed near-infrared reddening and the column density of the dust clouds. In this paper we particularly assess the feasibility of deriving the mass function of GMCs from near-infrared color excess data.}
   {We perform Monte Carlo radiative transfer simulations with 3D models of stellar radiation and clumpy dust distributions. We include the scattered light in the models and calculate near-infrared color maps from the simulated data. The color maps are compared with the true line-of-sight density distributions of the models. We extract clumps from the color maps and compare the observed mass function to the true mass function.}
   {For the physical configuration chosen in this study, essentially a face-on geometry, the observed mass function is a non-trivial function of the true mass function with a large number of parameters affecting its exact form. The dynamical range of the observed mass function is confined to $\sim 10^{3.5}\dots 10^{5.5}$ M$_\odot$ regardless of the dynamical range of the true mass function. The color maps are more sensitive in detecting the high-mass end of the mass function, and on average the masses of clouds are underestimated by a factor of $\sim 10$ depending on the parameters describing the dust distribution. A significant fraction of clouds is expected to remain undetected at all masses. }
   {The simulations show that the cloud mass function derived from $JHK$ color excess data using simple foreground screening geometry cannot be regarded as a one-to-one tracer of the underlying mass function.} 

   \maketitle
%

\section{Introduction}


The formation of GMCs is one of the major unsolved problems of the interstellar medium.
Studies of nearby Galactic GMCs provide information about their physical and chemical properties with a resolution that resolves the small-scale structure of the clouds. These studies have created a very detailed understanding of local GMCs such as the Orion complex, but an overall description of GMCs in the Galaxy remains incomplete due to confusion in the Galactic plane where most GMCs exist. In contrast to the GMCs of the Milky Way, nearby external galaxies provide an ``outside-view'' on the general structure and distribution of GMCs and therefore offer grounds for more meaningful characterization of these clouds, their environment, and their role in galaxy evolution.


Direct imaging of galaxies from the optical to near-infared offers an interesting, arcsecond-scale view of the dust distribution inside them. At these wavelengths the spatial resolution is limited by seeing, and sub-arcsecond resolution is easily achieved with standard imaging observations.
Qualitatively, the presence and distribution of GMCs can be traced by color maps made from observations in two broadband filters, e.g. $B-V$ or $J-K$. This approach has been exploited in a few studies of the dust content and GMCs of nearby galaxies (e.g. Trewhella \cite{trewhella98}; Howk \& Savage \cite{howk99}, \cite{howk00}; Regan \cite{regan00}; Thompson et al. \cite{thompson04}). Recent studies utilizing high signal-to-noise extinction mapping have indeed shown that mapping of galaxies in the near-infrared can provide a very detailed picture of their dust distribution, revealing structures that remain undetected in interferometric observations of CO due to lower sensitivity (Bialetski et al. \cite{bialetski05}).


It is, however, far from trivial to interpret the observed features of dust clouds in terms of physical quantities such as column density. The major difficulty arises from the combined effect of two factors: first, the relative geometry of dust clouds and stars along the line of sight is unknown. The dust clouds are embedded inside the galaxy at an unknown depth and are not between the galaxy and the observer as in the ``foreground screen geometry''. When a fraction of stars is in front of the cloud, the observed reddening of light is \emph{reduced} compared to the foreground screen geometry. Second, the observed flux contains not only the attenuated flux from background sources, but also the flux that is coming in from all directions and scattered towards the observer. As the scattering is stronger in shorter wavelengths, it results in a ``bluing'' of the observed colors, again \emph{reducing} the observed reddening. As a result of these two effects, the relation between the observed reddening, attenuation of surface brightness, and the actual column density cannot be easily assessed.


Motivated by the recent observational work of Bialetski et al. (\cite{bialetski05}) and Alves et al. (in prep.), who derive the observed mass functions of GMCs in galaxies using NIR reddening, we performed a radiative transfer simulation study to examine the feasibility of the method. In this paper we present the first results, addressing particularly the question of whether the form of the mass function can be reliably estimated using near-infrared color maps.


\section{Modeling}
\label{sec_modeling}
\subsection{Radiative transfer code}

We used the 3D Monte Carlo code developed by Juvela \& Padoan (\cite{juvela03}) and Juvela (\cite{juvela05}) to compute the flux of the scattered light. The code is essentially the same as used by Juvela et al. (\cite{juvela06}) to examine the scattered light in \emph{Galactic} dust clouds. For this study the code was altered with the possibility of creating photon packets from each computational cell to account for the stellar light of the galaxy. The relevant inputs for the program are the number density of each cell, the flux spontaneously emitted from each cell, and the parameters describing the dust properties.

In each run, a number of photon packages is created for each cell and wavelength, and the scattered flux, including multiple scatterings, is registered towards a selected direction.  
The final scattered flux is calculated as an average of several individual runs. The emitted flux and scattered flux are summed up to produce a map of total intensity. We use models that are $256^3$ cells in size. The output maps are computed so that their pixel size equals the cell size of the model; the intensity maps are thus $256^2$ pixels in size. To bind the model into physical units, the cell size of the model is fixed to 7 pc, which corresponds to the projected pixel size of a typical NIR detector at the distance of 5 Mpc (i.e. 0.3$''/$px).

We calculated the fluxes at the central wavelengths of $J$, $H$, and $K$  bands. The dust properties were based on those of Draine (\cite{draine03}) and we used the tabulated values for Milky Way and R$_V=3.1$, which are available on the web\footnote{http://www.astro.princeton.edu/$\sim$draine/dust/}. For the scattering function the Henyey-Greenstein scattering function is used (Henyey \& Greenstein \cite{henyey41}).

\subsection{Models}


The relative structure of dust and stars is undoubtly the most significant factor when considering the reddening effects. 
In this work we examine a simplified slab for which the distribution of stellar light is symmetric in the $xy$-plane. The light distribution in z-direction (line of sight) is described by an exponential, having a scale height $z_0^\mathrm{stars}$:
\begin{equation}
L_i = e^{-z/z_{0,i}^\mathrm{stars}}.
\end{equation}
Even though we neglect the gradients produced by spiral arms and the central bulge, this distribution can be regarded as the simplest possible model of a face-on spiral galaxy.
The same scale height is used for all near-infrared bands, $i=J$, $H$, $K$. Poissonian fluctuations are added to each cell, which result in 3-5 \% surface brightness fluctuations in the final maps. This is roughly the level expected for a bright face-on galaxy at the distance of 5 Mpc. Finally, the intensities at each band are scaled so that the models have intrinsic, i.e. unreddened, colors of $(J-H)_0 = 0.55$ and $(H-K)_0 = 0.24$. These colors are similar to what is observed in the interarm (supposedly unreddened) regions of spiral galaxies (e.g. Moriondo et al. \cite{moriondo01}).


The density distribution of each model is created by generating non-uniform cloudlets with statistical properties following given distributions. Spatially these clouds are placed randomly across the model volume following an exponential probability distribution parallel to the luminosity gradient and centered on the midplane of the model. 
The distribution of the center points of clouds is characterized by the scale height $z_0^\mathrm{clouds}$:
\begin{equation}
p = e^{-z/z_{0}^\mathrm{clouds}}.
\end{equation}
As the forms of both stellar light distribution and cloud distribution are similar, the ratio of scale heights, $\xi=z_0^\mathrm{clouds}$ / $z_0^\mathrm{stars}$, is the only parameter describing the relation between stars and clouds. Thus, the selection of numerical values for $z_0^\mathrm{clouds}$ and $z_0^\mathrm{stars}$ is of little importance. We use the scale height $z_0^\mathrm{stars}=210$ pc, which would represent the distribution of stars in the thin disk component of a galaxy (e.g. Wainscoat et al. \cite{wainscoat89}). With this setting of $z_0^\mathrm{stars}$, the models extend to 4.3 $\times z_0^\mathrm{stars}$ along the line of sight, enclosing 98.6 \% of the total luminosity. The remaining fraction is accounted for by creating photon packets isotropically at the edges of the model. The studies of photometric properties of edge-on galaxies have yielded estimates of ratio $\xi$ that vary between 20...75 \% (e.g. Wainscoat et al. \cite{wainscoat89}; Alton et al. \cite{alton00}). Regan et al. (\cite{regan95}) modelled the dust features in NGC 1530 and found $\xi=0.1\dots1.0$ depending on the position in the galaxy. However, the model used in Regan et al. is significantly different from ours. In the simulations we used the values $\xi=0.27$ and $0.5$. 

The original mass function of all generated clouds is selected to be a power law $dN/dM\sim M^{-\alpha_\mathrm{true}}$ with $\alpha_\mathrm{true}=1.5\dots 3$. The clouds are generated by first determining their masses from the original mass function and then distributing the mass around the center point of the cloud by selecting a random cell and inputing a fraction of clouds total mass into it. This is repeated until the cloud mass reaches the predetermined total mass. The cells are randomized so that the average radial density distribution of a cloud is $\rho \propto (r/r_c)^{-2}$ (Alves et al. \cite{alves98}), where the parameter $r_c$ is a random number between 1 and 10 cells, allowing clouds with similar masses to have differing radii. The total masses and positions of the clouds are recorded to perform comparisons with the observed cloud distribution.  
 The low-mass cut off of the input mass function, $M_\mathrm{low}$, is $\sim 10^5$ M$_\odot$ for most of the models. With the selected distance and pixel scale setting, a spherical cloud with $M = 10^5$ M$_\odot$, $r=10$ px, and $\rho\sim r^{-2}$ has an optical depth of $\tau_\mathrm{V}=3^m$ through its center. Below this mass, the observed reddening caused by clouds gets smaller than the error in color maps. Also, less massive clouds only extend across a few cells and are likely to be missed by the clump detection algorithm (see Sec. \ref{sec_undetected} and Fig. \ref{fig_completeness}). However, models with lower $M_\mathrm{low}$ were made to examine if the presence of diffuse clouds (which are present in the model but not resolved as individual clouds) affects the outcome. Likewise, some models were made with higher $M_\mathrm{low}$.

\begin{figure}
   \centering

   \includegraphics[width=0.65\columnwidth]{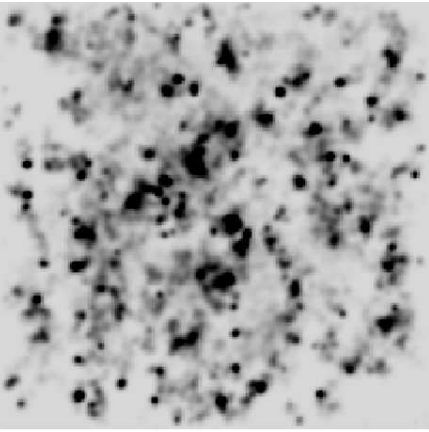}

      \caption{An example of column density distribution for model number \#1 (See Table \ref{tab_results}). }
         \label{fig_models}
   \end{figure}

The volume-filling factor of clouds in the midplane of the model is between 5\dots15 \%. This is set by differing the total number of clouds in the models. Thus, the total mass of the model is not fixed but depends on the number of input clouds.  For example, to create a model with a filling factor of 10 \%, about $\sim 200$ clouds must be created. These clouds contain $\sim 10^7 $ M$_\odot$, and their sizes vary from a few to a few hundred pixels. Figure \ref{fig_models} shows the distribution of column density for a typical model seen face-on.

The mass function of the clouds is a random realization from the true mass function. To generate a number of clouds, which is comparable to the amount that would be detected in a nearby face-on spiral galaxy with a similar clumpfind procedure, 10-50 different models are created for each set of input parameters. This results in $\sim$ 1500 clouds for each parameter set. The total mass in these clouds is on the order of $1\times 10^9$ M$_\odot$, a plausible value for the total gaseous mass of a spiral galaxy (e.g. Lundgren et al. \cite{lundgren04}).

\subsection{Simulated observations}


The $H-K$ and $J-H$ color maps are computed from the simulated maps at $JHK$ bands. Additional noise of 3-8\% is added to account for observational uncertainty, and maps are smoothed with a Gaussian kernel having a FWHM=1$''$ corresponding to the typical seeing. The color-excess method NICER (Lombardi \& Alves \cite{lombardi01}) is used to compute maps of visual extinction from the color maps. In NICER, the observed $H-K$ and $J-H$ colors of each pixel are compared to the colors of unreddened pixels, i.e. a reference region that is assumed to be free from extinction. From each model, a circular region with $r=5$ px is selected as the reference region.
 Applying NICER to all pixels of the observed $H-K$ and $J-H$ maps results in corresponding extinction map. 


Individual clouds are identified from the extinction map using the clumpfind routine ``clfind2d'' (Williams et al. \cite{williams94}). To be recognized by clfind2d, a cloud must have its extinction more than 2-$\sigma$ over the detection limit in all its pixels and the peak extinction must be over 5-$\sigma$. The clouds that are smaller than 0.5$''$ in diameter, i.e. half of the FWHM of the Gaussian that was used to smooth the color maps, are also rejected. The masses of extracted clouds are calculated by summing up extinction of all the on-cloud pixels, assuming foregound screening geometry and using $R_\mathrm{V}=3.1$ and the Milky Way ratio of $N(H_2)/A_\mathrm{V}=9.4\times 10^{20}$ cm$^{-2}$mag$^{-1}$ (Bohlin et al. \cite{bohlin78}).

\section{Results}        
\label{sec_results}

In the following we describe the main results concerning the derivation of mass function. A comprehensive description of results will be given in a forthcoming paper. 

\subsection{The cloud mass functions}

\begin{figure}
   \centering
    \includegraphics[width=\columnwidth]{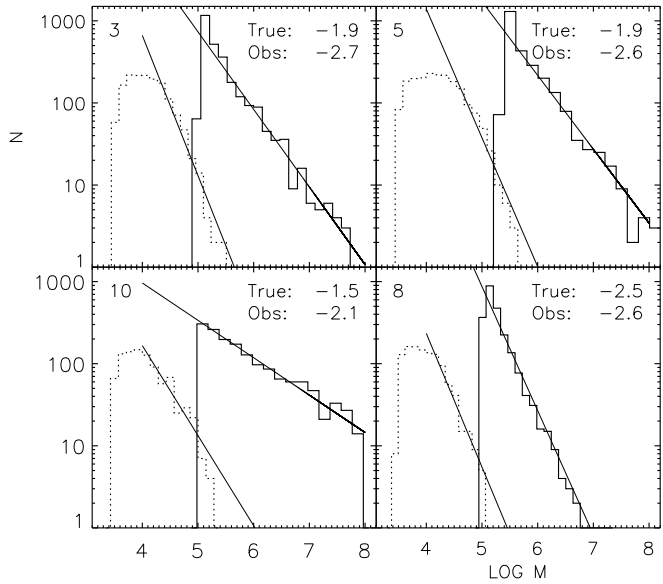}
      \caption{True (solid) and observed (dashed) mass functions of four models. $\alpha_\mathrm{true}$ and $\alpha_\mathrm{obs}$ are given in the frames. The numbering refers to the model numbers given in Table \ref{tab_results}.}
         \label{fig_cmf}
   \end{figure}

The cloud mass functions (MFs) derived for each model from simulated observations are compared with the MFs derived from the input column density maps with the same clumpfind procedure as described above. These ``true'' cloud MFs, which result from projecting the actual input mass distributions in two dimensions, were checked for consistency with the actual input MFs. Thus neither the projection effects nor overlapping of clouds alter the slopes of the mass function. Table \ref{tab_results} summarizes the results for 10 models. The histograms of true and observed MFs of four models are shown in Fig. \ref{fig_cmf} as an example.

The comparison of true and observed MFs shows two distinctive features. First, the observed MF spans a relatively small range in mass compared to the true MF. Regardless of the range of true MF, the observed one is always confined to roughly $10^{3.5}$\dots$10^{5.5}$ M$_\odot$. In particular, the high-mass end of the distribution is always terminated at $\sim10^{5.5}$. The MF peaks usually around $\sim 10^{4.2}$, turning over or flattening at lower masses. The constancy of the observed mass range also means that the total mass content of observed clouds is underestimated with a factor on the order of $\sim 10$.

Secondly, observed MFs are described well with power laws. Values of $\alpha_\mathrm{obs}$ span the range $\alpha_\mathrm{obs}=-2.1\dots-2.7$ differing significantly from $\alpha_\mathrm{true}$ in many cases. For a majority of models, $\alpha_\mathrm{obs}$ is significantly higher than $\alpha_\mathrm{true}$. Examples can be seen in Fig. \ref{fig_cmf}: the model \#3 with $\alpha_\mathrm{true}=-1.9$ is observed with $\alpha_\mathrm{obs}=-2.7$ (this is the case with the largest deviation), while some models with the same $\alpha_\mathrm{true}$ seem to recover the slope (e.g. \#8). Especially the models with relatively flat true MFs ($\alpha_\mathrm{true}=-1.5$) are observed to have steeper slopes (models \#9 and \#10). The observed slope is always steeper than the true slope, with the exception of model \#7 (and \#8) where the true slope is already very steep.

\begin{table}
\begin{minipage}[t]{\columnwidth}
\caption{Basic parameters of 10 models}            
\label{tab_results}     
\centering                         
\renewcommand{\footnoterule}{} 
\begin{tabular}{c c c c c c}      
\hline\hline               
\# & $z_0^\mathrm{clouds}$ / $z_0^\mathrm{stars}$ & $M_\mathrm{low}$\footnote{The low-mass cut off of the true MF} & $\alpha_\mathrm{true}$ & $\alpha_\mathrm{obs}$ & $\alpha_\mathrm{true} - \alpha_\mathrm{obs}$ \\   
\hline                      
   1  & 0.27 &  5.6 & -1.9 & -2.2 &  0.3        \\    
   2  & ``   &  5.2 & -1.8 & -2.3 &  0.5        \\    
   3  & ``   &  5.0 & -1.9 & -2.7 &  0.8        \\    
   4  & ``   &  5.0 & -2.0 & -2.3 &  0.3        \\    
   5  & 0.5  &  5.4 & -1.9 & -2.6 &  0.7        \\    
   6  & 0.5  &  5.0 & -1.9 & -2.4 &  0.5        \\    
   7  & 0.27 &  5.4 & -2.7 & -2.7 &    0        \\    
   8  & ``   &  5.0 & -2.5 & -2.6 &  0.1        \\    
   9  & ``   &  4.0 & -1.5 & -2.1 &  0.6        \\    
   10 & ``   &  5.0 & -1.5 & -2.1 &  0.6        \\    

\hline                             
\end{tabular}
\end{minipage}
\end{table}

\subsection{The fraction of undetected clouds}       
\label{sec_undetected}

\begin{figure}
   \centering

   \includegraphics[width=0.95\columnwidth]{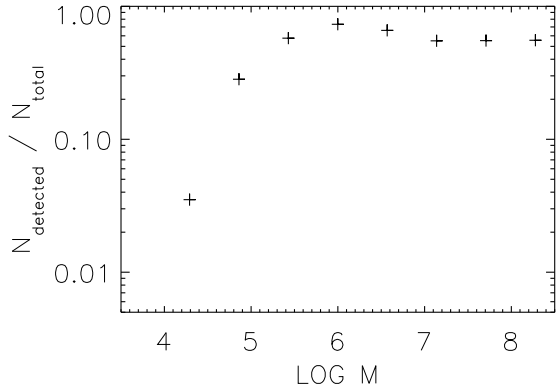}

      \caption{The fraction of detected clouds for different true masses. x-axis is the mass in solar masses, y-axis is $N_\mathrm{detected}/N_\mathrm{total}$.}
         \label{fig_completeness}
   \end{figure}

In addition to the underestimation of mass in the detected clouds, a fraction of the clouds remain undetected and thus their mass is completely missed. We calculated the fraction of detected clouds in each true mass bin by comparing the input column density maps to the derived color excess maps. Clumps that were detected with clfind2d with 2-$\sigma$ level as the lowest contour level were flagged as detections. The positions of detected clumps were then compared with the positions of input clumps and if the positions overlapped the particular clump was regarded as 'detected'. We summed up the detections of the models that have the same geometry. Figure \ref{fig_completeness} shows the resulting completeness curve for model \#9. The y-axis gives the number of detected clouds divided by the total number of clouds in the mass bin.


The fraction of detected clouds goes down quickly when the mass drops below $\sim 10^5$ solar masses. At $10^4$ M$_\odot$, only $\sim$1\% of the clouds are detected. The curve remains fairly constant for $M>10^{6}$, where about 60-80\% of clouds are detected. It should be emphasized that the x-axis of the diagram gives the \emph{original} masses of the clouds that are detected. The \emph{observed} masses of the clouds are significantly lower (as pointed out in Fig. \ref{fig_cmf}). The fraction that remains undetected at high masses results from the clouds that are located on the distant side of the galaxy and do not produce significant reddening features to the color maps.

\section{Discussion}    %
\label{sec_discussion}


The exact outcome of the simulation ultimately depends on a large number of parameters, the most important being the geometry of the model. Likewise, the recovered MFs are complicated reproductions of the true MFs and cannot be regarded simply as distributions that are ``shifted'' from higher to lower masses. 

Even though the true cloud masses can span several orders of magnitudes, the resulting MF is truncated to a modest dynamical range of approximately $10^{3.5}\dots10^{5.5}$ M$_\odot$. The truncation of the observed MF also sets an apparent upper limit for extinction that is typical of a given geometry. Derived from NIR observations, this upper limit is $\sim 2.5$ mags in units of $A_\mathrm{V}$, or correspondingly $10^{5.5}\dots10^6$ M$_\odot$ for a galaxy at 5 Mpc distance. This ``saturation'' of the observed colors has been also explored in earlier studies (Witt \& Gordon \cite{witt96}, \cite{witt00}; Regan \cite{regan00}). The saturation of the colors is a direct result of foreground emission and scattered light and it leads to the underestimation of masses seen in Fig. \ref{fig_cmf}. 

Although the power-law nature that is observed in galaxies is reproduced by the simulations, the observed slopes are systematically different from the true slopes. From the simulations presented in this paper it is not possible to describe how the slope is exactly affected, i.e. what the convolution is from the true mass function to the observed one. However, some qualitative remarks can be made immediately. There is an obvious trend that observed slopes are steeper than the true ones. The effect is stronger for flatter true slopes: models with $\alpha_\mathrm{true}\sim-1.5$ are observed with slopes of $\alpha_\mathrm{obs}=-2.1$, and models with $\alpha_\mathrm{true}\sim -2$ are observed with slopes  $\alpha_\mathrm{obs}=-2.2\dots 2.7$. The steepest models with $\alpha_\mathrm{true}= -2.7$ and -2.5 are observed to have almost the same slopes, so it would appear that very steep true mass functions are altered the least. From the parameters defining the dust distribution, i.e. $\xi$, filling factor, and $M_\mathrm{low}$, the first two are expected to have an effect to the observed mass function. The effect of $M_\mathrm{low}$ is supposedly small, as it mainly contributes to the diffuse dust distribution that is not resolved as individual clouds. Thus it affects the determination of background colors, but has a neglible effect on the clouds that are detected. We will present a more precise discussion about the effect of these parameters in a forthcoming paper.


It should be emphasized that the treatment in this study neglects some, possibly severe, sources of confusion, such as embedded star clusters and uncertainties in the clump definition. The effect of simplifications is that the overall variance in the observed mass functions is unavoidably underestimated. Even with the simplistic case considered in this study the correspondence between observed and underlying true mass functions is not unique.

Despite the complexity in the interpretation of color excess data, the use of color maps to trace GMCs remains interesting due to its superb resolution and the low noise of the column density maps. This is confirmed by the simple observation that the color-excess maps indeed trace the morphology of faint dust structures that remain undetected in interferometric line observations. The use of NIR color maps can possibly be complemented with observations in other wavelengths and/or with radiative transfer modeling to improve the estimate of the underlying mass function. However, determination of physical masses and mass function relying merely on the NIR color data should be taken with caution.

\begin{acknowledgements}
We thank the referee, S. Bianchi, for several comments that improved the paper significantly.
\end{acknowledgements}

\end{document}